	\documentclass[
		a4paper,american,USenglish,authorcolumns,
		numberwithinsect,
		anonymous=false%
	]{lipics-v2021}
	\def\docclass{lipics}
    \usepackage[dvipsnames]{xcolor}
	\def\version{arxiv}

	\def\draftmode{false}  

\usepackage[utf8]{inputenc}
\makeatletter
\usepackage{xifthen}

\newcommand\iflipics[2]{\ifthenelse{\equal{\docclass}{lipics}}{#1}{#2}}
\newcommand\ifkoma[2]{\ifthenelse{\equal{\docclass}{koma}}{#1}{#2}}
\newcommand\ifieee[2]{\ifthenelse{\equal{\docclass}{ieee}}{#1}{#2}}
\newcommand\ifsiam[2]{\ifthenelse{\equal{\docclass}{siam}}{#1}{#2}}
\newcommand\ifsiamsingle[2]{\ifthenelse{\equal{\docclass}{siam-single}}{#1}{#2}}
\newcommand\ifmysiam[2]{\ifthenelse{\equal{\docclass}{my-siam}}{#1}{#2}}
\newcommand\ifacm[2]{\ifthenelse{\equal{\docclass}{acm}}{#1}{#2}}
\newcommand\ifdcc[2]{\ifthenelse{\equal{\docclass}{dcc}}{#1}{#2}}
\newcommand\ifspringerjournal[2]{\ifthenelse{\equal{\docclass}{springer-journal}}{#1}{#2}}
\newcommand\iflncs[2]{\ifthenelse{\equal{\docclass}{lncs}}{#1}{#2}}
\ifthenelse{ \equal{\docclass}{lipics} \OR \equal{\docclass}{koma} \OR \equal{\docclass}{ieee} \OR \equal{\docclass}{siam} \OR \equal{\docclass}{my-siam} \OR \equal{\docclass}{acm} \OR \equal{\docclass}{dcc} \OR \equal{\docclass}{springer-journal} \OR \equal{\docclass}{lncs} }{
	\PackageInfo{paper}{Building paper with docclass = \docclass} 
}{
	\PackageWarning{paper}{docclass = "\docclass", but must be one of "lipics", "koma", "ieee", "siam", "siam-single", "my-siam", "acm", "dcc", "springer-journal", "lncs"}
}

\newcommand\ifmanuscript[2]{\ifthenelse{\equal{\version}{manuscript}}{#1}{#2}}
\newcommand\ifarxiv[2]{\ifthenelse{\equal{\version}{arxiv}}{#1}{#2}}
\newcommand\ifsubmission[2]{\ifthenelse{\equal{\version}{submission}}{#1}{#2}}
\newcommand\ifproceedings[2]{\ifthenelse{\equal{\version}{proceedings}}{#1}{#2}}
\ifthenelse{ 
	\equal{\version}{manuscript} 
	\OR \equal{\version}{arxiv} 
	\OR \equal{\version}{submission} 
	\OR \equal{\version}{proceedings} 
}{
	\PackageInfo{paper}{Building paper version = \version} 
}{
	\PackageWarning{paper}{version = "\version", but must be one of "manuscript", "arxiv", "submission", "proceedings"}
}

\newcommand\ifdraft[2]{\ifthenelse{\equal{\draftmode}{true}}{#1}{#2}}
\ifthenelse{ \equal{\draftmode}{true} \OR \equal{\draftmode}{false} }{
	\PackageInfo{paper}{Building paper with draftmode = \draftmode} 
}{
	\PackageWarning{paper}{draftmode = "\draftmode", but must be "true" or "false"}
}

\usepackage[T1]{fontenc}
\ifsiam{
	\usepackage{lmodern}
	\usepackage{slantsc}
}{}
\ifsiamsingle{
	\usepackage{lmodern}
	\usepackage{slantsc}
}{}
\ifmysiam{
	\usepackage{lmodern}
	\usepackage{slantsc}
}{}
\ifkoma{
	\usepackage{lmodern}
	\usepackage{slantsc}
}{}
\iflipics{
	\usepackage{lmodern}
	\usepackage{slantsc}
}{}
\ifdcc{
	\usepackage{lmodern}
	\usepackage{slantsc}
}{}
\ifspringerjournal{
	\usepackage{lmodern}
	\usepackage{slantsc}
	\usepackage{xcolor}
}{}
\iflncs{
	\usepackage{lmodern}
	\usepackage{slantsc}
	\usepackage{xcolor}
}{}

\DeclareFontShape{T1}{lmr}{b}{sc}{<->ssub*cmr/bx/sc}{}
\DeclareFontShape{T1}{lmr}{bx}{sc}{<->ssub*cmr/bx/sc}{}

\usepackage{babel}
\input{ushyphex.tex} 

\usepackage{array,multicol,multirow}
\ifieee{
	\usepackage[cmex10]{amsmath,mathtools}
	\usepackage{amsfonts,amssymb}
}{}
\ifkoma{
	\usepackage{amsmath,amsfonts,amssymb,mathtools}
}{}
\iflipics{
	\usepackage{amsmath,amsfonts,amssymb,mathtools}
}{}
\ifsiam{
	\usepackage{amsmath,amsfonts,amssymb,mathtools}
}{}
\ifsiamsingle{
	\usepackage{amsmath,amsfonts,amssymb,mathtools}
}{}
\ifmysiam{
	\usepackage{amsmath,amsfonts,amssymb,mathtools}
}{}
\ifacm{
	\usepackage{mathtools}
}{}
\ifdcc{
	\usepackage{amsmath,amsfonts,amssymb,mathtools}
}{}
\ifspringerjournal{
	\usepackage{amsmath,amsfonts,amssymb,mathtools}
}{}
\iflncs{
	\usepackage{amsmath,amsfonts,amssymb,mathtools}
}{}

\usepackage{mleftright}\mleftright 
\usepackage{relsize,xspace,booktabs,adjustbox,needspace,pbox,relsize,makecell}
\ifieee{
		
		\usepackage{enumitem}
}{
	\ifspringerjournal{
		\usepackage{enumitem}
		\setlist{topsep=\medskipamount}
	}{
		\usepackage{enumitem}
	}
}
\usepackage{graphicx}

\ifacm{}{
	\usepackage{colonequals}
}

\usepackage{wref}

\ifsiam{
	\usepackage{ltexpprt}
}{}
\ifsiamsingle{
	\usepackage{ltexpprt}
}{}

\newdimen\makeboxdimen
\newcommand\makeboxlike[3][l]{%
\setbox0=\hbox{#2}%
\global\makeboxdimen=\wd0%
\setbox1=\hbox{\makebox[\makeboxdimen][#1]{%
\makebox[0pt][#1]{#3}%
}}%
\ht1=\ht0%
\dp1=\dp0%
\box1%
}

\newcommand\plaincenter[1]{%
	\mbox{}\hfill#1\hfill\mbox{}%
}

\ifthenelse{\equal{\docclass}{koma} \OR \equal{\docclass}{my-siam}}{
	\setlength\parindent{1.5em}
	\usepackage[headsepline]{scrlayer-scrpage}
	\pagestyle{scrheadings}
	\clearscrheadfoot
	\AtBeginDocument{%
		\automark[section]{}%
	}
	\ohead{\pagemark}
	\rehead{\mytitle}
	\lohead{\headmark}
	\addtokomafont{caption}{\sffamily\small}
	\addtokomafont{captionlabel}{\sffamily\textbf}
	\setcapmargin{2em}
}{}
\ifmysiam{
	\setcapmargin{1em}
	\setcapindent{0em}
}{}
\ifmysiam{
	\setlength\parskip{0pt}
	\RedeclareSectionCommand[
		beforeskip=-1.25\baselineskip,
		afterskip=0.75\baselineskip,
	]{section}
	\RedeclareSectionCommand[
		beforeskip=-1\baselineskip,
		afterskip=-1.5em,
	]{subsection}
	\RedeclareSectionCommand[
		beforeskip=-1\baselineskip,
		afterskip=-1.5em,
	]{subsubsection}
	\RedeclareSectionCommand[
		beforeskip=-.25\baselineskip,
		indent=1.5em,
		afterskip=-1em,
	]{paragraph}
}{}
\ifdcc{
	\ifproceedings{}{
		\pagestyle{plain}
		\setlength{\footskip}{5ex}
	}
}{}

\AtBeginDocument{%
	\let\mytitle\@title%
}

\ifsiam{
	\usepackage[bibtex-alpha]{url-doi-arxiv}
}{}
\ifsiamsingle{
	\usepackage[bibtex-alpha]{url-doi-arxiv}
}{}
\ifmysiam{
	\usepackage[bibtex-alpha]{url-doi-arxiv}
}{}
\ifkoma{
	\usepackage[bibtex-alpha]{url-doi-arxiv}
}{}
\iflipics{
	\usepackage[bibtex]{url-doi-arxiv}
}{}
\ifdcc{
	\usepackage[bibtex-alpha]{url-doi-arxiv}
}{}
\ifspringerjournal{
	\usepackage[bibtex-alpha]{url-doi-arxiv}
}{}
\iflncs{
	\usepackage[bibtex-alpha]{url-doi-arxiv}
}{}

\let\oldthebibliography\thebibliography
\renewcommand\thebibliography[1]{%
	\oldthebibliography{#1}%
	\pdfbookmark[1]{References}{}%
}

\usepackage{lscape} 

\ifkoma{
	\usepackage{newfloat}
	\DeclareFloatingEnvironment[%
			name=Algorithm,%
			placement=thb,%
		]{algorithm}
}{}
\iflipics{
	\usepackage{newfloat}
	\DeclareFloatingEnvironment[%
			name=Algorithm,%
			placement=thb,%
		]{algorithm}
}{}
\ifmysiam{
	\usepackage{newfloat}
	\DeclareFloatingEnvironment[%
			name=Algorithm,%
			placement=thb,%
		]{algorithm}
}

\ifdcc{
	\usepackage{float}
	\usepackage[font={small},labelfont=bf]{caption}
}{}
\ifmysiam{
	\setcounter{topnumber}{3}
	\setcounter{bottomnumber}{3}
	\setcounter{totalnumber}{3}     
	\setcounter{dbltopnumber}{3}    

}{}
\ifsiam{

	\setcounter{topnumber}{3}
	\setcounter{bottomnumber}{3}
	\setcounter{totalnumber}{3}     
	\setcounter{dbltopnumber}{3}    

}{}
\ifsiamsingle{

	\setcounter{topnumber}{3}
	\setcounter{bottomnumber}{3}
	\setcounter{totalnumber}{3}     

}{}

\usepackage{dcolumn}

\usepackage{wclrscode}

\usepackage{textcomp} 
\usepackage{listings}

\usepackage{tikz}

\usetikzlibrary{positioning,arrows.meta,fit}
\usetikzlibrary{backgrounds,calc,trees,graphs}
\usetikzlibrary{shapes.geometric,shapes.misc}
\usetikzlibrary{decorations,decorations.pathreplacing}
\usetikzlibrary{scopes,patterns,bending}

\pgfdeclarelayer{background}
\pgfsetlayers{background,main}

\usetikzlibrary{external}
\tikzsetexternalprefix{pics/externalized/}
\tikzset{
	external/system call={%
		lualatex \tikzexternalcheckshellescape -halt-on-error %
			-interaction=batchmode -jobname "\image" "\texsource"%
	},
}
\tikzset{external/export=false} 

\iflipics{
	\newtheorem{fact}[theorem]{Fact}

	\newenvironment{proofof}[1]{%
		\begin{proof}[{{Proof of #1{}}}]%
	}{%
		\end{proof}%
	}
}{}
\ifacm{
	\AtEndPreamble{
		\theoremstyle{acmdefinition}
		\newtheorem{remark}[theorem]{Remark}
		\newtheorem{fact}[theorem]{Fact}
	}
	
	\newenvironment{proofof}[1]{%
		\begin{proof}[{{Proof of #1{}}}]%
	}{%
		\end{proof}%
	}
}{}
\ifsiam{
	\newtheorem{remark}{Remark}
	\newenvironment{proofof}[1]{%
		\begin{proof}[{{#1{}}}]%
	}{%
		\end{proof}%
	}
}{}
\ifsiamsingle{
	\newtheorem{remark}{Remark}
	\newenvironment{proofof}[1]{%
		\begin{proof}[{{#1{}}}]%
	}{%
		\end{proof}%
	}
}{}
\iflncs{
	\spnewtheorem{fact}[theorem]{Fact}{\itshape}{}
	
	\let\orig@endproof\endproof
	\def\endproof{\qed\orig@endproof}
	\newenvironment{proofof}[1]{%
		\begin{proof}[{{#1{}}}]%
	}{%
		\end{proof}%
	}
}{}
\ifthenelse{%
		\equal{\docclass}{lipics} \OR \equal{\docclass}{siam} \OR 
		\equal{\docclass}{siam-single} \OR \equal{\docclass}{acm} \OR
		\equal{\docclass}{lncs}%
}{}{
	\usepackage[amsmath,hyperref,thmmarks]{ntheorem}
	
	\theoremheaderfont{\sffamily\upshape\bfseries}
	\theorembodyfont{\slshape}
	\theoremseparator{:}
	\newtheoremstyle{proofstyle}%
	  {\item[\theorem@headerfont\hskip\labelsep ##1\theorem@separator]}%
	  {\item[\theorem@headerfont\hskip\labelsep ##3\theorem@separator]}
	
	\theorempreskip{\topsep} 

	\theoremsymbol{\adjustbox{scale=.8}{$\triangleleft\mkern-1mu$}}
	
	\newtheorem{theorem}{Theorem}[section]
	
	\theoremstyle{plain}
	\theorempreskip{\topsep}

	\newtheorem{definition}[theorem]{Definition}
	
	\theoremstyle{plain}
	\theorembodyfont{\upshape}

	\theoremsymbol{\raisebox{-.25ex}{$\Box$}}
	\qedsymbol{\raisebox{-.25ex}{$\Box$}}
	
	\theoremstyle{proofstyle}
	\theoremheaderfont{\sffamily\slshape}
	\newtheorem{proof}{Proof}

}

\iflipics{
		\newenvironment{thmenumerate}[2][]{%
			\begin{enumerate}[
				label={\textsf{\textbf{\color{darkgray}{\makebox[\widthof{(a)}][c]{\textup{(\alph*)}}}}}},
				ref={\ref{#2}\kern.1em--\kern.1em(\alph*)},
				itemsep=0pt,
				topsep=.5ex,
				leftmargin=1.75em,
				#1
			]%
		}{%
			\end{enumerate}%
		}
}{
	\ifspringerjournal{
		\newenvironment{thmenumerate}[2][]{%
			\begin{enumerate}[
				label={\makebox[\widthof{(a)}][c]{\textup{(\alph*)}}},
				ref={\ref{#2}\kern.1em--\kern.1em(\alph*)},
				itemsep=0pt,
				topsep=\smallskipamount,
				leftmargin=1.75em,
				#1
			]%
		}{%
			\end{enumerate}%
		}
	}{
		
	}
}

\newcommand*\ie{\mbox{i.\hspace{.2ex}e.}}
\newcommand*\eg{\mbox{e.\hspace{.2ex}g.}}

\newcommand*\wrt{\mbox{w.\hspace{.2ex}r.\hspace{.2ex}t.}\xspace}

\newcommand\Z{\mathbb Z}

\usepackage{fixmath}

\newcommand{\ESymbol}{\mathbb{E}}

\newcommand{\ProbSymbol}{\ensuremath{\mathbb{P}}}

\DeclarePairedDelimiterXPP\Prob[1]{\ProbSymbol}[]{}{%
	#1%
}
\DeclarePairedDelimiterXPP\E[1]{\ESymbol}[]{}{%
	#1%
}
\DeclarePairedDelimiterXPP\Eover[2]{\ESymbol_{#1}}[]{}{%
	#2%
}
\DeclarePairedDelimiterXPP\ProbIn[2]{\ProbSymbol_{#1}}[]{}{%
	#2%
}
\providecommand{\Prob}{} 
\providecommand{\ProbIn}{} 
\providecommand{\E}{} 
\providecommand{\Eover}{} 

\iflipics{}{
	\makeatletter
	\let\oldalign\align
	\let\endoldalign\endalign
	\renewenvironment{align}{%
		\begingroup%
		\let\oldhalign\halign
		\def\halign{%
			\let\oldbreak\\%
			\def\nonnumberbreak{\nonumber\oldbreak*}%
			\def\\{%
				\@ifstar{\nonnumberbreak}{\oldbreak}%
			}%
			\oldhalign%
		}
		\oldalign%
	}{%
		\endoldalign%
		\endgroup%
	}
}
\newcommand*\numberthis[1][]{\stepcounter{equation}\tag{\theequation}}

\allowdisplaybreaks[3]

\newcommand\splitaftercomma[1]{%
  \begingroup
  \begingroup\lccode`~=`, \lowercase{\endgroup
    \edef~{\mathchar\the\mathcode`, \penalty0 \noexpand\hspace{0pt plus .25em}}%
  }\mathcode`,="8000 #1%
  \endgroup
}

\def\mydots{\xleaders\hbox to.5em{\hfill.\hfill}\hfill}
\newlength\tmpLenNotations

\ifdraft{%
	\iflipics{
		\usepackage{lineno} 
	}{
		\usepackage[switch]{lineno} 
	}
	\linenumbers
	\overfullrule=6mm
	
	\usepackage[color,notref,notcite]{showkeys}
	\definecolor{refkey}{gray}{.99}
	\colorlet{labelkey}{green!60!black!60}
	
	\usepackage[inline,nolabel]{showlabels}

	\showlabels{cite}
	\showlabels{citealt}
	\showlabels{citealp}
	\showlabels{citet}
	\showlabels{Citet}
	\showlabels{citep}
	\showlabels{citeauthor}
	\showlabels{Citeauthor}
	\showlabels{citefullauthor}
	\showlabels{citeyear}
	\showlabels{citeyearpar}
	\showlabels{wref}
	\showlabels{wpref}
	\showlabels{wtpref}
	\showlabels{wildref}
	\showlabels{wildpageref}
	\showlabels{wildtpageref}
}{}

\iflipics{
	\ifmanuscript{\hideLIPIcs}{}
	\ifarxiv{\hideLIPIcs}{}
	\ifsubmission{}{\nolinenumbers}
}{}

\ifdraft{}{%
	\usepackage{microtype}
}

\hypersetup{
	final,
	unicode=true, 
	bookmarks=true,
	bookmarksnumbered=true,
	bookmarksdepth=2,
	bookmarksopen=true,
	breaklinks=true,
	hidelinks,
}

\newsavebox\tmpbox

\ifdcc{
	\renewcommand\paragraph{\@startsection{paragraph}{4}{\parindent}
	                                      {\smallskipamount}
	                                      {-1em}%
	                                      {\normalfont\normalsize\bfseries}}
}{}
\iflipics{
	\let\oldparagraph\paragraph
	\renewcommand\paragraph[1]{%
		\subparagraph*{#1.}
	}
}{
	\let\oldparagraph\paragraph
	\renewcommand\paragraph[1]{%
		\oldparagraph{#1.}
	}
}

\ifmysiam{
	\let\oldsubsection\subsection
	\renewcommand\subsection[1]{%
		\oldsubsection{#1.}%
	}
	\let\oldsubsubsection\subsubsection
	\renewcommand\subsubsection[1]{%
		\oldsubsubsection{#1.}%
	}
}{}
\ifsiam{
	\let\oldsubsection\subsection
	\renewcommand\subsection[1]{%
		\oldsubsection{#1.}%
	}
	\let\oldsubsubsection\subsubsection
	\renewcommand\subsubsection[1]{%
		\oldsubsubsection{#1.}%
	}
}{}
\ifsiamsingle{
	\let\oldsubsection\subsection
	\renewcommand\subsection[1]{%
		\oldsubsection{#1.}%
	}
	\let\oldsubsubsection\subsubsection
	\renewcommand\subsubsection[1]{%
		\oldsubsubsection{#1.}%
	}
}{}

\let\epsilon\varepsilon

\def\myacknowledgements{}
\ifkoma{
	\newcommand\acknowledgements[1]{\def\myacknowledgements{\paragraph{Acknowledgements}#1}}
}{}
\ifieee{
	\newcommand\acknowledgements[1]{\def\myacknowledgements{\section*{Acknowledgement}#1}}
}{}
\ifsiam{
	\newcommand\acknowledgements[1]{\def\myacknowledgements{\section*{Acknowledgement}#1}}
}{}
\ifsiamsingle{
	\newcommand\acknowledgements[1]{\def\myacknowledgements{\section*{Acknowledgement}#1}}
}{}
\ifmysiam{
	\newcommand\acknowledgements[1]{\def\myacknowledgements{\section*{Acknowledgement}#1}}
}{}
\ifdcc{
	\newcommand\acknowledgements[1]{\def\myacknowledgements{\section*{Acknowledgement}#1}}
}{}
\ifacm{
	\newcommand\acknowledgements[1]{\def\myacknowledgements{
		\section*{Acknowledgement}#1%
	}}
}{}
\ifspringerjournal{
	\newcommand\acknowledgements[1]{\def\myacknowledgements{\bmhead{Acknowledgements}#1}}
}{}
\iflncs{
	\newcommand\acknowledgements[1]{\def\myacknowledgements{%
		\begin{credits}
		\subsubsection{\ackname} #1
		\end{credits}
	}}
}{}

\setlist[description]{font=\boldmath}

\makeatother

\usepackage{diagbox}

\usepackage{hyperref}

\ifacm{
		\title[Short Title]{My Long Paper Title}
}{}
\ifspringerjournal{
	\title[Short Title]{My Long Paper Title}
}{
	\title{Virtual-Memory Powersort}
}

\iflipics{
	\author{Finn Moltmann}{Philipps-Universität Marburg, Germany}{moltmann@uni-marburg.de}{https://orcid.org/0009-0009-0367-0521}{}
	
	\author{Tamio-Vesa Nakajima}{Philipps-Universität Marburg, Germany}{nakajima@informatik.uni-marburg.de}{https://orcid.org/0000-0003-3684-9412}{}
	
	\author{Sebastian Wild}{Philipps-Universität Marburg, Germany and University of Liverpool, UK}{wild@informatik.uni-marburg.de}{https://orcid.org/0000-0002-6061-9177}{Supported by EPSRC grant EP/X039447/2.}

	\authorrunning{F.~Moltmann, T.-V.~Nakajima and S.~Wild}
	\Copyright{Finn Moltmann, Tamio-Vesa Nakajima and Sebastian Wild}


    \supplementdetails[linktext={Virtual Memory Powersort Implementation}, swhid={swh:1:snp:8e65d4e51fc7e84ac78dde26a19bea76773d8ce8;origin=https://git.sr.ht/~tamionv/vm-powersort}]{Software}{https://archive.softwareheritage.org/swh:1:snp:8e65d4e51fc7e84ac78dde26a19bea76773d8ce8;origin=https://git.sr.ht/~tamionv/vm-powersort}
}{}

\acknowledgements{Note: Where we were aware of both a conference and a full paper, we cite both to give credit to both the definitive version and the venue of first dissemination.}

\usepackage{bm}

\nolinenumbers{}
\begin{document}

\ifacm{}{\maketitle} %

\begin{abstract}
We give a more space-efficient implementation of adaptive mergesort: Virtual-Memory Powersort.
Using internal buffering techniques, we significantly reduce the memory consumption of the algorithm; specifically, for sorting $n$ objects the required buffer area is reduced from space for $n/2$ objects to $O(\sqrt{n \log n})$ objects. 
While this space-efficiency can be achieved (indeed reduced to $O(1)$) conceptually very easily with known inplace merging algorithms, using these as a drop-in replacement for the standard merge algorithm incurs a substantial slow-down.
Virtual-Memory Powersort, by contrast, uses the \textit{same} number of moves and comparisons as previous Powersort implementations up to an additive $O(n)$ term. We report on an empirical running-time study comparing our implementation against other Powersort variants and state-of-the-art stable sorting methods, demonstrating that \emph{almost} in-place stable sorting can be achieved with negligible overhead in many scenarios.
\end{abstract}

%
%
%
%

%
%
%
%
%
%
%
%
%

%
%
%
%
%
%
%
%


\section{Introduction}

Sorting is an elementary building block that most standard libraries offer an implementation of.
For general purpose methods, the specification often prescribes a \emph{stable} sorting method,
\ie, one where elements that compare equal \wrt the sorting criteria remain in the same relative order as
in the input.
Apart from stability, the desiderata for a library sort are to be fast and to use little (or no) extra memory.
Unfortunately, these three properties seem to be at odds: 
while Quicksort is usually the fastest generic option when implemented \emph{in place},%
\footnote{%
	Strictly speaking, ``in place'' should mean no extra memory at all;
	usually $O(\log n)$ (words of) space are accepted as ``in place''.
	In practice, a stack of $O(\log n)$ pointers~-- as recursion stack in Quicksort and top-down Mergesort, or as run stack in Timsort/Powersort~--  has negligible memory footprint and is indeed is often implemented as a static fixed-size array.
} 
it is not stable. 
Mergesort is easily implemented as a stable method and can be made almost as fast as Quicksort, but typically at the expense of a linear-size buffer of extra memory.
In-place variants of Mergesort are known, but they either sacrifice stability~\cite{Kronrod1969,KatajainenPasanenTeuhola1996,Reinhardt1992,EdelkampWeiss2014,Wild2018a,EdelkampWeissWild2020} or are substantially more complicated and slower in practice~\cite{McFadden2021,Franceschini2007}.

\begin{figure}[tbp]
	\colorlet{buffercolor}{black!10}
	\colorlet{runacolor}{RoyalBlue}
	\colorlet{runbcolor}{lipicsYellow}
	\sffamily
	
	\pgfdeclarepatterninherentlycolored{mergeresult}
		{\pgfpointorigin}{\pgfpoint{8pt}{8pt}}
		{\pgfpoint{8pt}{8pt}}
		{
			\pgfpathrectangle{\pgfpoint{0pt}{0pt}}{\pgfpoint{4pt}{8pt}}
			\pgfsetfillcolor{runacolor}
			\pgfusepath{fill}
			\pgfpathrectangle{\pgfpoint{4pt}{0pt}}{\pgfpoint{8pt}{8pt}}
			\pgfsetfillcolor{runbcolor}
			\pgfusepath{fill}
		}

	\small
	
	\plaincenter{
		\mbox{\makeboxlike[l]{\textbf{\textsf{Step 2:}}}{\textbf{\textsf{Step 1:}}}}%
		\qquad
			\begin{scriptsize}
				\begin{tikzpicture}[scale = 0.55,semithick]
				\draw[fill=runacolor] (0,0) -| ++(3,1) -- cycle;
				\draw[fill=runbcolor] (3,0) -| ++(3,1) -- cycle;
				\draw[fill=buffercolor] (7,0) rectangle ++(3,1) ;
				\draw[-{Stealth[bend]}](1.5,0.5) ..controls(1.5,2) and (8.5, 2.1 ).. node[above] {swap} (8.5,1);
				\draw(3,0) -- (3,1);
				\end{tikzpicture}
			\end{scriptsize}		
	}
	
	\smallskip
	
	\plaincenter{
		\mbox{\makeboxlike[l]{\textbf{\textsf{Step 2:}}}{\textbf{\textsf{Step 2:}}}}%
		\qquad
			\begin{scriptsize}
				\begin{tikzpicture}[scale = 0.55,semithick]
			\draw[fill=buffercolor] (0,0) rectangle (3,1) ;
			\draw[fill=runbcolor] (3,0) -| ++(3,1) -- cycle ;
			\draw[fill=runacolor] (7,0) -| ++(3,1) -- cycle ;
			\draw[{Stealth[bend]}-] (1.5,1) to[in=180,out=90,looseness=1] (3,1.6);
			\draw (3,1.6) to[out=0,in=90,looseness=.6] node[above=-1pt,pos=.3] {merge} (8.5,.5);
			\draw  (3,1.6) to[out=0,in=90,looseness=1] (4.5,.5);
				\end{tikzpicture}
			\end{scriptsize}
	}
	
	\bigskip
	\medskip
	
	\plaincenter{
		\mbox{\makeboxlike[l]{\textbf{\textsf{Step 2:}}}{\textbf{\textsf{Result:}}}}%
		\qquad
			\begin{scriptsize}
			\begin{tikzpicture}[scale = 0.55,semithick]
				\draw[pattern=mergeresult] (0,0) -| ++(6,1) -- cycle;
				\draw[fill=buffercolor] (7,0) rectangle ++(3,1);
			\end{tikzpicture}
		\end{scriptsize}
	}
	
	\caption{%
		Standard ``copy''-merging procedure where one of the two runs fits into the buffer as used in CPython's Powersort implementation.
		(Buffer and input need not be adjacent in memory.)
	}
	\label{fig:merge-copy}
\end{figure}
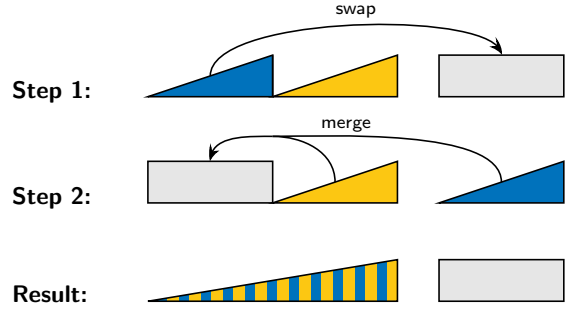

Orthogonally to the triad of stability, time, and space, 
standard libraries have increasingly embraced the idea of \emph{adaptive sorting}, 
where the sorting process becomes faster when the input is already partially ordered.
Asymptotically optimal algorithms have been devised for many thinkable measures of presortedness~\cite{EstivillCastro1992,PeterssonMoffat1992,PeterssonMoffat1995},
but not always is the overhead worth the potential savings~\cite{ElmasryHammad2009}.
One exception has been exploiting natural \emph{runs}, \ie, contiguous sorted segments in the input.
This approach has found wide-spread use via Timsort~\cite{Peters2002} and is used for stable sorting 
in most popular software frameworks (CPython, PyPy, OpenJDK, Android Runtime, Swift, \textellipsis).
Several of these frameworks have replaced Timsort's original merge policy by \emph{Powersort}~\cite{MunroWild2018,CawleyGellingNebelSmithWild2023}.
All these Timsort/Powersort implementations still use extra space for $n/2$ objects
to facility efficient stable merging; see \wref{fig:merge-copy}.
This limits the use of these successful adaptive stable methods to use cases where ample extra
space is available.

\begin{figure}[tbp]
	\medskip
    \centering
    \sffamily\small
    \begin{tikzpicture}[semithick,>=stealth,scale=.9]

    \begin{scope}

    \begin{scope}

    \begin{scope}[local bounding box=freedA]
    
    \node[overlay] at (2.75,-.4) {\textbf{Run 1}};
    
    \begin{scope}[local bounding box=A1]
    \draw[gray] (0, 0) -- (1, 0) -- (1, 0.5) -- cycle;
    \end{scope}
    
    \begin{scope}[xshift=1.5cm, local bounding box=A2]
    \draw[gray] (0, 0) -- (1, 0) -- (1, 1) -- (0, .5) -- cycle;
    \end{scope}
    
    \end{scope}

    \draw[very thick, red, shorten >= -.3cm, shorten <= -.3cm] (freedA.north east) -- (freedA.south west);
    \draw[very thick, red, shorten >= -.3cm, shorten <= -.3cm] (freedA.north west) -- (freedA.south east);
    
    \draw[red, very thick, ->, shorten >=.1cm] (3.5, 2) -- (3.5, 1.25);
    \begin{scope}[xshift=3cm, local bounding box=A3]
    \begin{scope}
    \clip (0, 0) -- (1, 0) -- (1, 1.5) -- (0, 1) -- cycle;
    \filldraw[fill=RoyalBlue, draw=black] (.5, -.5) -- (1, -.5) -- (1, 2) -- (.5, 2) -- cycle;
    \end{scope}
    \draw[gray] (0, 0) -- (.5, 0) -- (.5, 1.25) -- (0, 1) -- cycle;
    \draw[black] (.5, 0) -- (1, 0) -- (1, 1.5) -- (.5, 1.25) -- cycle;
    \end{scope}
    
    \begin{scope}[xshift=4.5cm, local bounding box=A4]
    \filldraw[fill=RoyalBlue, draw=black] (0, 0) -- (1, 0) -- (1, 2) -- (0, 1.5) -- cycle;
    \end{scope}

    \end{scope}
    
    \draw[gray, ->, shorten <=.1cm, shorten >= .1cm] (A1.east) -- (A2.west);
    \draw[,gray, ->, shorten <=.1cm, shorten >= .1cm] (A2.east) -- (A3.west);
    \draw[->, shorten <=.1cm, shorten >= .1cm] (A3.east) -- (A4.west);
    
    \end{scope}
    
    \begin{scope}[xshift=6.5cm]
	
	\node[overlay] at (2.75,-.4) {\textbf{Run 2}};
    \begin{scope}
    
    \begin{scope}[local bounding box=A1]
    \draw[gray] (0, 0) -- (1, 0) -- (1, 0.5) -- cycle;
    \end{scope}
    
    \draw[very thick, red, shorten >= -.3cm, shorten <= -.3cm] (A1.20) -- (A1.200);
    \draw[very thick, red, shorten >= -.3cm, shorten <= -.3cm] (A1.160) -- (A1.-20);
    
    \draw[red, very thick, ->, shorten >=.1cm] (2, 2) -- (2, .75);

    \begin{scope}[xshift=1.5cm, local bounding box=A2]
    \begin{scope}
    \clip (0, 0) -- (1, 0) -- (1, 1) -- (0, .5) -- cycle;
    \filldraw[fill=lipicsYellow, draw=black] (.5, -.5) -- (1, -.5) -- (1, 2) -- (.5, 2) -- cycle;
    \end{scope}
    \draw[gray] (0, 0) -- (.5, 0) -- (.5, .75) -- (0, .5) -- cycle;
    \draw[black] (.5, 0) -- (1, 0) -- (1, 1) -- (.5, .75) -- cycle;
    \end{scope}
    
    \begin{scope}[xshift=3cm, local bounding box=A3]
    \filldraw[fill=lipicsYellow, draw=black] (0, 0) -- (1, 0) -- (1, 1.5) -- (0, 1) -- cycle;
    \end{scope}
    
    \begin{scope}[xshift=4.5cm, local bounding box=A4]
    \filldraw[fill=lipicsYellow, draw=black] (0, 0) -- (1, 0) -- (1, 2) -- (0, 1.5) -- cycle;
    \end{scope}

    \end{scope}
    
    \draw[gray, ->, shorten <=.1cm, shorten >= .1cm] (A1.east) -- (A2.west);
    \draw[->, shorten <=.1cm, shorten >= .1cm] (A2.east) -- (A3.west);
    \draw[->, shorten <=.1cm, shorten >= .1cm] (A3.east) -- (A4.west);
    \end{scope}
    
    \begin{scope}[yshift=-3.75cm, xshift=2.25cm]
		\node at (3.5,-.4) {\textbf{Output Run }};
    \begin{scope}
    
    \begin{scope}[local bounding box=A1]
    \begin{scope}
    \clip (0, 0) -- (1, 0) -- (1, 0.5) -- cycle;
    \fill[pattern=mergeresult] (-.5, -.5) -- (1, -.5) -- (1, 2.5) -- (-.5, 2.5) -- cycle;
    \end{scope}
    \draw (0, 0) -- (1, 0) -- (1, 0.5) -- cycle;
    \end{scope}
    
    \begin{scope}[xshift=1.5cm, local bounding box=A2]
    \begin{scope}
    \clip (0, 0) -- (1, 0) -- (1, 1) -- (0, 0.5) -- cycle;
    \fill[pattern=mergeresult] (-.5, -.5) -- (1, -.5) -- (1, 2.5) -- (-.5, 2.5) -- cycle;
    \end{scope}
    \draw (0, 0) -- (1, 0) -- (1, 1) -- (0, 0.5) -- cycle;
    \end{scope}

    \begin{scope}[xshift=3cm, local bounding box=A3]
    \begin{scope}
    \clip (0, 0) -- (1, 0) -- (1, 1.5) -- (0, 1) -- cycle;
    \fill[pattern=mergeresult] (-.5, -.5) -- (1, -.5) -- (1, 2.5) -- (-.5, 2.5) -- cycle;
    \end{scope}
    \draw (0, 0) -- (1, 0) -- (1, 1.5) -- (0, 1) -- cycle;
    \end{scope}
    
    \begin{scope}[xshift=4.5cm, local bounding box=A4]
    \begin{scope}
    \clip (0, 0) -- (1, 0) -- (1, 2) -- (0, 1.5) -- cycle;
    \fill[pattern=mergeresult] (-.5, -.5) -- (1, -.5) -- (1, 2.5) -- (-.5, 2.5) -- cycle;
    \end{scope}
    \draw (0, 0) -- (1, 0) -- (1, 2) -- (0, 1.5) -- cycle;
    \end{scope}
    
    \begin{scope}[xshift=6cm, local bounding box=A5]
    \draw (0, 0) -- (1, 0) -- (1, 2.5) -- (0, 2) -- cycle;
    \end{scope}
    
    \end{scope}
    
    \draw[red, very thick, ->, shorten >=.1cm] (6, 2.75) -- (6, 2);
    \draw[->, shorten <=.1cm, shorten >= .1cm] (A1.east) -- (A2.west);
    \draw[->, shorten <=.1cm, shorten >= .1cm] (A2.east) -- (A3.west);
    \draw[->, shorten <=.1cm, shorten >= .1cm] (A3.east) -- (A4.west);
    \draw[->, shorten <=.1cm, shorten >= .1cm] (A4.east) -- (A5.west);
    \end{scope}
    
    \end{tikzpicture}
    
	\caption{%
		Merging two runs in virtual memory.  Both input runs and the output run are contiguous in virtual address space, but physically broken into \emph{pages} which need not be contiguous in physical memory.
		Whenever a page is entirely consumed form an input run (grayed/crossed-out in the figure), it is released to be reused as a new output page for further merging.
	}
	\label{fig:paged-merge}
\end{figure}
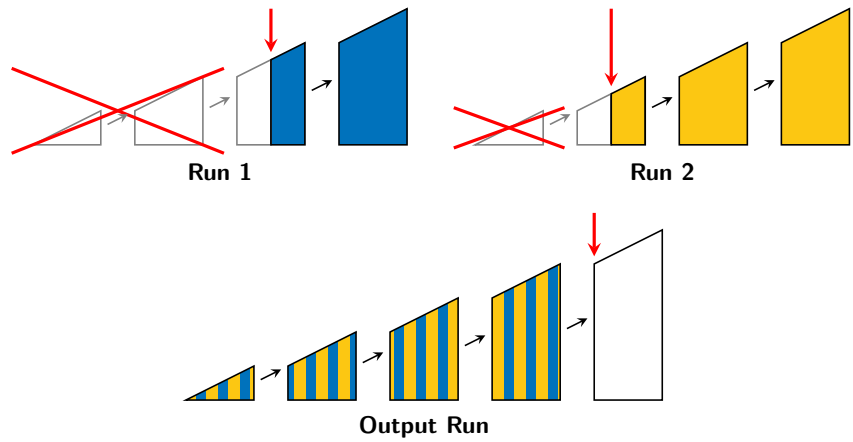

We show that a seasoned technique, \emph{virtual memory},%
\footnote{%
	The algorithmic idea is described, \eg, as ``block table'' by  \cite[\S5.3]{WittenMoffatBell1999} in the context of external sorting. 
	We borrow the name from the ``virtual runs'' of Graefe~\cite[\S3.4]{Graefe2006} from sorting in database systems.
}
can be efficiently combined with Powersort to 
yield $O(\sqrt{n\log n})$-extra-space variants with almost no running-time overhead 
over the copy-based version as used in CPython.
The key idea of virtual memory is to break the physical memory (our input array) into logical \emph{pages}
(each holding $P$ objects) and to represent contiguous runs by (potentially non-contiguous) 
sequences of pages.
It is a folklore result that merging and Mergesort can be implemented (\eg, in external memory) by reading from and writing to only one block of data from the input resp.\ output runs.
We apply the same method at page granularity. Moreover, whenever a page of an input run has been exhausted, we declare that memory area a \emph{free page}, to be reused as a future output page (see \wref{fig:paged-merge}).
The novelty of our results lies in making this efficient for \emph{internal sorting}.

Merging page by page, yields the output as a \emph{virtual run}: sorted in virtual order, but with pages potentially in arbitrary locations.
A permutation step, swapping contents of entire pages, can produce the run as a contiguous block in physical memory and thus turn the virtual-memory merge into a conventional one; 
however, this permuting step is rather costly.
For the final merge this extra cost is inevitable, but for \emph{intermediate} merges we can leave pages permuted: If each page stores the physical location of its successor page, we can sequentially traverse a virtual run almost as efficiently as a run contiguous in physical memory.

Note that by itself, the virtual-memory technique does not yield a fully inplace algorithm, 
since we need a small number of extra pages to bootstrap the process.
Moreover, we need extra space for one pointer per page to store successor pages.
In theory, at most 3 extra page are sufficient to handle merges, if we allow runs to start and end mid-page; that however complicates the merging logic substantially.
Our second insight is that for Powersort specifically, we can ensure that each written run starts at a page boundary 
(and ends in a potentially partially used page) if we have space for $O(\log n)$ additional pages: 
at any point in time, we only have $O(\log n)$ written runs on the run stack of Powersort.
These are the only runs that use partial pages; 
the not-yet-discovered runs in the suffix of the input do not yet start at a page boundary.%
\footnote{%
    We note that the virtual-memory technique is otherwise not specific to Powersort in any way,
    and could likely be applied to other stack-based adaptive mergesort variants.
}

Using loop unrolling and a careful implementation of memory accesses, our VM Powersort implementation executes, for most of its operation, exactly the same code as standard Powersort.
Indeed, since we typically do \emph{fewer} data movements in VM Powersort, 
when sorting large objects (that are not too expensive to compare), VM Powersort can be \emph{faster} than a CPython-style Powersort.

Our closer study of Powersort's merging choreography is of interest for another reason: 
When allowed a size-$n$ buffer, 
we can implement all but the last merges as an \emph{out-of-place merge}, reading data from one area and writing the output to a different area, avoiding any extraneous copies. (The CPython implementation does such extraneous copies for every merge).
For non-adaptive mergesort, this is a well-known folklore trick sometimes called ``Ping-Pong Merge''~\cite{ChandramouliGoldstein2014}, where alternating levels of Mergesort's recursion tree switch the roles of two arrays (read from $A$, write to $B$, resp.\ vice versa). 
For run-adaptive Mergesort, the ``level'' of a merge may not be known a priori and hence this trick becomes harder to execute. 
The specific merge policy of Powersort, however, allows for a \emph{Pingpong Powersort} implementation that avoids all extraneous copies of elements up to \emph{one} move per element.

\subsection*{Related Work}

Making sorting use little extra space has long been of interest, especially in combination with stable sorting.
A comprehensive survey is thus not possible here and we restrict ourselves to key results for mergesort.
For a pedagogical overview of various inplace merging methods, as well as other inplace sorting algorithms, see Wegner's lecture notes~\cite{Wegner2014}.

Interestingly, one can obtain an inplace sorting method without making merging itself inplace;
this works via a technique of internal buffer, using part of the input as buffer for merges.
This can be exploited via unbalanced merging~\cite{Kronrod1969,KatajainenPasanenTeuhola1996}
or via quicksort-style partitioning~\cite{Reinhardt1992,EdelkampWeiss2014,Wild2018a,EdelkampWeissWild2020};
neither of these approaches is \emph{stable}, though.
Already these methods are classified as ``not efficient in practice'' by Katajainen et al.~\cite{KatajainenPasanenTeuhola1996}.

A line of work starting with Kronrod~\cite{Kronrod1969} designed merge methods without extra space.
The first resp.\ more practical ones were not stable~\cite{MannilaUkkonen1984,HuangLangston1988}, but 
various \emph{stable} inplace merging methods have been devised, as well~\cite{Pardo1977,HuangLangston1992}.
While these methods remained of mostly theoretical interest,
the simple divide-and-conquer based merge method of Dudzinski and Dydek~\cite{DudzinskiDydek1981}
is the basis of the inplace merge method implemented as part of the GNU Compiler Collection's STL implementation (more details in \wref{sec:algorithms});
theoretically speaking its running time of $\Theta(n\log n)$ for merging two runs of size $n$
is suboptimal, though.
Later refinements of Dudzinski and Dydek's method~\cite{KimKutzner2004,KimKutzner2008,GeffertKatajainenPasanen2000} improve upon that but are, again, more complicated.

In theory, one can combine stable and inplace merging even with a linear number of moves~\cite{Franceschini2007}, but the complicated resulting method is likely unpractical and has not been implemented to our knowledge.
The most competitive available practical implementation of inplace, stable sorting may be Wikisort~\cite{McFadden2021}, which is based on Kim and Kutzner's refinement~\cite{KimKutzner2008}.


\section{Algorithms}
\label{sec:algorithms}

Both Powersort and Timsort are stable, run-adaptive variants of Mergesort. 
They operate by one outer iteration over the array from left to right while maintaining a stack of runs.
In each step, they find the next run of already sorted data to be put on top of the stack. 
A certain set of rules specific to the algorithm, the \emph{merge policy}, determines whether 
we execute (potentially several) merges of runs at the top of stack before adding the newly detected run.
Timsort's original merge policy uses a combination of rules comparing certain lengths among the 4 topmost runs.
Powersort's merge policy~\cite{MunroWild2018} enforces increasing \emph{``(run-boundary) powers''} (defined below) on the stack, simulating Mehlhorn's nearly optimal binary search tree algorithm~\cite{Mehlhorn1977,Mehlhorn1984}. 
This more robust merge policy achieves optimal adaptivity up to an additive linear term with very low running-time overhead.

Throughout, we let $T$ denote the size, in words, of the data type we are
sorting. We always count memory in words. By \emph{moves} we mean moving one
element of the data type we are sorting to a different location in memory.

\begin{algorithm}
        \begin{codebox}
            \Procname{$\proc{Powersort}({\id{input}[0, n)})$}
            \zi \Comment{Adaptively sort $\id{input}[0, n)$.}
            \li $\id{work} \gets \proc{NewBuffer}(\id{input}[0, n))$.
            \li $\id{stack} \gets \text{a stack of pairs of buffers and powers}$.
            \li $\id{curr} \gets \proc{ExtractRun}(\id{work})$.
            \li \While $\lnot \proc{Empty}(\id{work})$ \Do
                \li $\id{next} \gets \proc{ExtractRun}(\id{work})$
                \li $p = \proc{Power}(n, \id{curr}, \id{next})$
                \li \While $\proc{Top}(\id{stack}) \isequal (\id{top}, p')$ and $p' \geq p$ \Do
                    \li $\id{curr} \gets \proc{MergeBuffers}(\id{top}, \id{curr})$
                    \li $\proc{Pop}(\id{stack})$
                \End
                \li $\proc{Push}(\id{stack}, (\id{curr}, p))$
                \li $\id{curr} \gets \id{next}$
            \End
            \li \While $\proc{Top}(\id{stack}) = (\id{top}, p)$ \Do
                \li $\id{curr} \gets \proc{MergeBuffers}(\id{top}, \id{curr})$
            \End
            \li $\proc{Copy}(\id{curr}, \id{input}[0, n))$.
        \end{codebox}
    \caption{Abstract Powersort}\label{alg:abstract}
\end{algorithm}

In this section, we detail the various algorithms we will use. Let us begin with
a very abstract version of Powersort --- this formulation will strip away
details that are not relevant to the current work, as well as be general enough
to capture all the variants of Powersort we will discuss. This algorithm is
given in \wref{alg:abstract}. Throughout this algorithm, we will assume some
mechanism to maintain a small number \emph{buffers} of bounded total length.
These buffers should act as though they were double ended queues
(e.g.~with the typical operations on such queues i.e.~\proc{PushFront}, \proc{PushBack}, \proc{PopFront}, \proc{PopBack}). We will not directly use the operations \proc{PushFront}, \proc{PushBack}, \proc{PopFront}, \proc{PopBack}, only the abstract operations \proc{NewBuffer}, \proc{ExtractRun}, \proc{MergeBuffers}, \proc{Copy} and \proc{Power}. The meanings of these operations will now be described:
\begin{description}
\def\proc#1{\textrm{\textsc{#1}}}
    \item[$\bm{\proc{NewBuffer}}(s)$.] Instantiate a buffer using the memory from array $s$; this operation lets our buffering scheme take ownership of $s$.
    \item[$\bm{\proc{ExtractRun}}(b)$.] Extracts a maximal non-decreasing run from $b$; in principle this can also use e.g.~insertion sort to insure a minimum run length.
    \item[$\bm{\proc{MergeBuffers}}(b, b')$.] Merge buffers $b$ and $b'$, assumed to be in non-decreasing order, into a result buffer. This may destroy $b$ and $b'$.
    \item[$\bm{\proc{Copy}}(b, s)$.] Copies the buffer $b$ into the array $s$. This assumes that $s$ was the input to a \proc{NewBuffer} earlier; this relinquishes ownership of $s$.
    \item[$\bm{Power}(n, b, b')$.] A function which assigns an integer to a pair of buffers $b, b'$. This will be described below; for now, assume it it chosen so as to make \wref{alg:abstract} efficient.
\end{description}

Our efficient implementations will reuse
memory to actually maintain these buffers --- however for \wref{alg:abstract}, the buffers simply behave like (double-ended) queues. 
Furthermore,
every buffer will always contain elements from a contiguous subsegment of the
original array --- for ease of presentation, we implicitly assume that each
buffer remembers this subsegment.\footnote{All segments are indexed closed on the left and open on the right.}

The main important fact about powersort is that the \emph{merge cost} is bounded
by the \emph{run-length entropy}. First let us formally define merge cost.

\begin{definition}
    For a merge-based sorting algorithm, the \emph{merge cost} is the sum of the
    lengths of all the inputs to all the merges used within the algorithm.
\end{definition}

Throughout, we let $\mathcal{H}$ denote the run-length entropy for our input
sequence~--- formally, for an input consisting of runs of lengths $\ell_1,\ldots,\ell_r$, we have $\mathcal H = \sum_{i=1}^r \frac{\ell_i}{n} \lg \frac{n}{\ell_i}$ 
(which coincides with the entropy of the distribution of selecting a random run proportional to its length).

\begin{theorem}[{{\cite[Thm 3.3, Rem 3.4]{CawleyGellingNebelSmithWild2023}}}]
    Consider an integer $n$ and two runs $a, b$, such that
    \[
        1 \leq \proc{Left}(a) \leq \proc{Right}(a) = \proc{Left}(b) \leq
        \proc{Right}(b) \leq n.
    \]
    
    \plaincenter
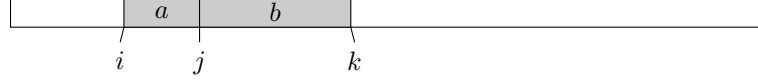
{
    	\begin{tikzpicture}[yscale=.4,xscale=.5]
    		\draw  (0,0) rectangle ++(20,1) ;
    		\draw[fill=black!20] (3,0) rectangle node {$a$} (5,1) ;
    		\draw[fill=black!20] (5,0) rectangle node {$b$} (9,1) ;
    		\draw (3,0) -- ++(-.1,-.5) node[below] {$i$} ; 
    		\draw (5,0) -- ++(0,-.5) node[below] {$j$} ; 
    		\draw (9,0) -- ++(.1,-.5) node[below] {$k$} ; 
    	\end{tikzpicture}
    }
    
    Suppose $i = \proc{Left}(a)$, $j = \proc{Right}(a) = \proc{Left}(b)$, $k =
    \proc{right}(b)$.
    If we set $\proc{Power}(n, a, b)$ to be the smallest integer $p$ such that
    there is a $c\in \Z$ with $\frac{i + j}{2n} < c\cdot 2^{-p} \leq \frac{j + k}{2n}$
    then, within algorithm \wref{alg:abstract} the following hold:
    \begin{enumerate}
        \item The stack has height bounded by $\lceil \log_2 n \rceil + 1$.
        \item The merge cost is bounded by $n(\mathcal{H} + 2)$.
    \end{enumerate}
\end{theorem}

We will focus on four different implementations of \wref{alg:abstract}, which
we detail below. For all that follows, let $\mathcal{M}$ denote the merge cost
of our algorithm. Note that many of our implementations will use $
\mathcal{M}+(n-1)$ comparsions --- this is because run detection takes an extra $n-1$
comparisons; for brevity, we drop the $-1$.

\begin{description}
    \item[CPython-style Powersort.] 
    	This is our baseline.
    	In this implementation, buffers are merely
        subsegments of the original memory, in the natural order.
        \proc{NewBuffer}, \proc{ExtractRun} and \proc{Copy} are purely metadata
        operations. Only \proc{Merge} remains, which (i) copies the smaller
        buffer to auxiliary storage, then (ii) merges the two buffers back into
        the original space they used. This approach uses, in the worst case,
        $\mathcal{M} + n$ comparisons and $1.5 \mathcal{M}$ moves, as well as $\frac n2\cdot T$ extra memory.
        
        Note that our code uses the same buffer-copy strategy as the CPython implementation of Powersort,
        but in other aspects (\eg, non-galloping merge) chooses
        what is efficient for a \texttt{C++} generic sort with moderately expensive comparisons.

    \item[Pingpong Powersort (new).] In this implementation, buffers on the
        stack are maintained in auxiliary memory, while \id{curr}, \id{next},
        \id{work} are left in the original storage. 
        Altogether, in the worst case we use $\mathcal{M}
        + n$ comparisons, $\mathcal{M} + n$ moves, and $nT$ extra memory. We will explain this approach in more detail in \wref{sec:fewMoves}.

    \item[Virtual-Memory (VM) Powersort (new).] In this implementation, we use a variant
        of internal buffering to eliminate most of the extra space usage of our
        algorithm. 
        We will explain this approach in detail in \wref{sec:almostInplace}; for
        now, we simply state that this approach will use $\mathcal{M} + n$
        comparisons, $\mathcal{M} + 3n$ moves, and $O(\sqrt{nT \log n})$ extra memory. 

    \item[Inplace-Merge Powersort.] In this implementation, we store all buffers
        within the original array, but achieve a space usage of $O(\log n)$ extra words of space 
        via the 
        inplace merge method \texttt{std::\_\_merge\_without\_buffer},%
        \footnote{
        	Source at:
        	\url{https://github.com/gcc-mirror/gcc/blob/master/libstdc\%2B\%2B-v3/include/bits/stl_algo.h\#L2436C5-L2436C27}
        }
        the (nonstandard) inplace merge implementation of the GCC STL implementation.
        It is based on the divide-and-conquer method of~\cite{DudzinskiDydek1981} and has running time $O(n\log n)$ for merging two runs of size $\Theta(n)$.
        Within the STL, it is used in \texttt{std::stable\_sort} in case no extra memory is available.
        
\end{description}

We close with an informal description of our new algorithms.
Pingpong Powersort uses the initial array $A[0..n)$ and 
an auxiliary array $B[0..n)$ as buffers: 
each run will be stored in $A$ or $B$, and we merge runs from $A$ and/or $B$ 
using the corresponding positions of $A$ and/or $B$ as working space. 

Virtual-memory Powersort, by contrast, treats the input array as the (first and largest) 
part of a collection of $n/P + \log_2(n)+O(1)$ \emph{pages} of $P$ elements each. 
(For the purposes of this paragraph, assume we sort objects 1 word long.)
Each run will occupy a (not necessarily contiguous) list of full pages and maybe one partially filled page 
(and needs to also store pointers to these pages). 
As of yet undiscovered runs lie in their original position in the input array, so they do not waste space;
only the $\log_2(n)+O(1)$ runs on the run stack each require one partially filled page, 
and we need $O(1)$ extra pages while merging.
The partial pages and pointers cost $O(n/P + P \log n)$ extra words of memory. 
Setting $P = \Theta(\sqrt{n / \log n})$ minimizes the extra memory requirement, 
yielding $O(\sqrt{n \log n})$.

\subsection{Galloping Merges}

CPython's Timsort/Powersort implementation uses a \emph{galloping merge}~\cite{Peters2002,McIlroy1993}, \ie, exponential searches for the minimum of the smaller array in the larger array.
Using such a merge method makes the \emph{comparison} cost of many run-adaptive mergesort variants
approach the \emph{dual}-runlength entropy~\cite{GhasemiJugeKhalighinejad2024,GhasemiJugeKhalighinejad2022}; however, the data movement remains mostly unchanged.

Since galloping merges increase the variability of running times and is mostly effective if comparisons are very expensive~\cite{MunroWild2018},
we do not include galloping in our merge methods.
It is, however, an interesting question for future work to evaluate the overhead of virtual memory for galloping, since on one hand a check for page boundaries needs to be added, on the other hand, there is an enticing novel opportunity for speedups through virtual memory: 
in case an entire page can be skipped over, we can make this page part of the output using only metadata changes, \emph{entirely avoiding} data movement for this page.


\section{Pingpong Powersort (Few moves)}\label{sec:fewMoves}

Let us first give a very simple implementation of \wref{alg:abstract} which sorts using $\mathcal{M} + n$ moves and $O(nT)$ extra memory. This implementation allocates a new buffer of $n$ elements, each of size~$T$, call it $\id{stackBuf}$. Consider some buffer $\id{buf}$, which contains elements from the range $\id{input}[i, j)$. (Recall that each buffer only holds a contiguous subrange of the original array.) If $\id{buf}$ is one of $\id{work}, \id{curr}, \id{next}$, then we keep it in the memory range $\id{input}[i, j)$. Otherwise (i.e.~if $\id{buf}$ is kept in $\id{stack}$), we keep it in the memory range $\id{stackBuf}[i, j)$.

Let us describe how realize the operations from \wref{alg:abstract} with this implementation.

\begin{description}
\def\proc#1{\textrm{\textsc{#1}}}
    \item[$\bm{\proc{NewBuffer}}(s)$.] This operation does nothing in this implementation: \id{work} ought already to be stored in \id{input}.
    \item[$\bm{\proc{ExtractRun}}(b)$.] This operation simply looks for the first pair of adjacent elements in the wrong order in \id{work}, then returns that. The result doesn't ever need to be moved (since it is stored either in \id{curr} or \id{next}).
    \item[$\bm{\proc{MergeBuffers}}(b, b')$.] We observe that we only ever merge a buffer from \id{stack} with \id{curr}. For some values of $i \leq j \leq k$, the first occupies the memory range $\id{stackBuf}[i, j)$ and the second the memory range $\id{input}[j, k)$. Our goal will be to merge these two ranges into $\id{input}[i, k)$~-- this can be done simply by merging left-to-right and outputting from~$i$ towards~$k$ in \id{input} (as in Step~2 of \wref{fig:merge-copy}). This takes at most $k - i$ moves i.e.~as many moves as elements in the output.
    \item[$\bm{\proc{Copy}}(b, s)$.] This operation does nothing: $\id{curr}$ is already in \id{input}.
\end{description}

A careful reader will have noticed that we have not yet described one detail: given the above description of where we store our buffers, we need to copy buffers from \id{input} to \id{stackBuf} whenever pushing it to the stack. However, this happens at most once for every element in the input array. Hence, as we do these $n$ extra moves and otherwise do exactly one move for each comparison, we see that we used at most $\mathcal{M} + n$ moves as required.

In fact, we can sometimes do slightly better. Suppose the stack is such that, in line 7 of \wref{alg:abstract}, we need to pop (and merge) at least one element. As described before, our algorithm extracts runs from the top of the stack, merging them with \id{curr}, storing them in memory range within \id{input}. However, at the very last pop, we can instead directly merge \id{curr} into \id{stackBuf}: this time we need merely to merge right-to-left. Any such run eliminates its unnecessary moves. (This seems to happen very often in our experiments, cf.~\wref{fig:assignments}.)

\section{Virtual-Memory Powersort (Almost Inplace)}
\label{sec:almostInplace}

In this section we will explain how to implement the buffers in
\wref{alg:abstract} so as to use only $O(\sqrt{nT \log n})$ extra memory, as well
as only $\mathcal{M} + 3n$ moves.
We will first give a simple theoretical algorithm that achieves these
constraints, and then optimise some implementation details in
\wref{sec:almostImpl}.

Our algorithm will manage memory in terms of \emph{pages} of $P$ words of memory each, where
we choose $P \approx \sqrt{n / T \log_2 n}$.%
\footnote{%
	For our practical
	implementation, we additionally chose $P$ to be a power of 2, to make arithmetic
	faster.
} 
Each buffer will be stored in a sequence of pages, where the contents of
a page are in the correct order, but pages may be located at arbitrary locations. Each page by itself will also be stored contiguously in memory. 
Thus each buffer will own a list of pointers to its constituent pages. 
Our algorithm globally maintains an (initially empty) list of free pages, 
which we call the \emph{free list}.
Whenever a new page is requested we try to take it from the free list. 
If no free page is available, we allocate a new page.

Let us now walk through the implementation of all operations:
\begin{description}
\def\proc#1{\textrm{\textsc{#1}}}
    \item[\proc{PushFront}, \proc{PushBack}.] We try to insert into the
        first/last page if possible; if not, we get a new page for the
        buffer from the free list, and insert the new element there.
    \item[\proc{PopFront}, \proc{PopBack}.] This operation is simple to
        implement; the only crucial detail is that if a full page becomes empty,
        we add it to the free list.
    \item[\proc{NewBuffer}.] This operation takes in memory not yet managed by
        our page system, and creates a buffer containing it. We handle this by
        segmenting the memory into pages.\footnote{There may be a final partial page --- this page will never be declared free and reused by our algorithm, so we may treat it simply as a normal page, even if we don't own all of it.}
    \item[\proc{Copy}.] This operation is quite nontrivial in our case, since we
        are copying from a buffer to the original memory~-- while our buffer
        may still be using part of the original memory! To handle this, we do
        the following: we search in the page list of the buffer \id{curr} to see if
        the first page in \id{input} is used somewhere. If it is, we copy that
        page out to a new page. Then, we copy the correct contents into the
        first page of \id{input}. We then continue this approach for all further
        pages.%
        \footnote{One page in particular will hold what
            should be the contents of the partial page that may exist at the end
            of \id{work}. This page is even easier to handle, since we know that
            the memory at the end of \id{work} is not being used to hold
            anything, as we do not free a partial page.}%
        ${}^{\mkern-1.4mu\text{,}\mkern0.8mu}$%
        \footnote{Note also that this theoretically uses quadratic time with respect to the number of pages, i.e. $O((N / P)^2)$ which for our choice is $O(n T \log_2 n)$. This can be optimised to theoretical $O(n)$ time easily by using e.g.~a hash table, but in practice a linear scan is fast enough.}
    \item[\proc{ExtractRun}, \proc{MergeRuns}.] These operations are implemented
        trivially in terms of the previous operations, i.e.~one allocates a
        result run and builds it by using \proc{PushBack}, while using
        \proc{PopFront} to consume the input buffer(s).
    \item[\proc{Empty}.] This is a trivial metadata operation.
\end{description}

Let us now consider the total memory consumption of this approach. To do this,
let us count how many pages we need at any point in the run of the
algorithm, and how many pages we receive from segmenting the original memory.
Furthermore, we need to count how much metadata memory we use.

\begin{description}
    \item[Pages needed.] Clearly there are at most $\lceil n / P \rceil$
        \emph{full} pages. Note that there are at most 2 partially
        full pages for every buffer. Furthermore, there are only $\lceil \log_2
        n\rceil$ buffers on the stack, as well as $O(1)$ extra auxilliary
        buffers needed at any point during the run of the algorithm. Hence
        overall we need $n / P + O(\log_2(n))$ pages.
    \item[Metadata needed.] For every buffer, we must keep a page list,
        containing pointers to each page within our buffer. Since this is stored
        as a linked list, each buffer requires $O(1)$ extra memory plus $O(1)$
        extra memory for each full page. Hence, since there are only $\log_2 n
        + O(1)$ buffers overall, and $\lceil n / P \rceil$ full pages, this only
        requires $n / P + \log_2 n + O(1)$ words of space overall.
    \item[Pages received.] From segmenting the original main memory, we gain
        $\lfloor n / P \rfloor = n / P - O(1)$ pages in the worst case.
\end{description}

Hence we need $O(n / P)$ extra words of metadata space, as well as $O(\log_2n)$
extra pages in the worst case, each costing $P\cdot T$ words, 
for $T$ the size of the sorted records in words. 
Setting 
$P \approx \sqrt{n / (T \log_2 n)}$ 
balances these two terms, for a final memory usage of
$O(\sqrt{nT\log n})$.

\subsection{Implementation considerations}\label{sec:almostImpl}

\paragraph{Fewer partial pages}
In order to simplify our implementation, we can observe that the only time we
ever use \proc{PopFront} is either on \id{work}, or when calling \proc{Merge}.
The practical relevance of this is that we can maintain as an invariant that all
buffers (except \id{work} and during \proc{Merge}) have all pages entirely full
\emph{except perhaps the last one}. This tightens the analysis on the number of extra
pages needed by a factor of 2, significantly lowering memory consumption, as
well as simplifying implementation logic.

\paragraph{Fast merging}
The innermost loop of our algorithm is a merge~-- hence it makes sense to
optimise the merging of two buffers as much as possible. If we were to naively
merge by calling \proc{PopFront} and \proc{PushBack}, we would incur the cost of
a second indirection at each step! Hence, we implement our algorithm \emph{page
by page}: while either input buffer has any contents left, we merge
the first pages of the two inputs into the currently last page of the output.

\begin{algorithm}
    \begin{minipage}{.5\linewidth}
        \begin{codebox}
            \Procname{$\proc{PageMerge}(a[x_a, y_a), b[x_b, y_b), r[x_r, y_r))$}
            \zi \Comment{Merge as much as possible of}
            \zi \Comment{$a[x_a, y_a)$ and $b[x_b,
            y_b)$ into $r[x_r, y_r)$.}
            \li \While $x_a < y_a \land x_b < y_b \land{} $\smash{\fbox{{$x_r < y_r$}}} \Do
                \li \If $a[x_a] \leq b[x_b]$ \Then
                    \li $r[x_r] \coloneqq a[x_a]$
                    \li $x_a \coloneqq x_a + 1$
                \li \Else
                    \li $r[x_r] \coloneqq b[x_b]$
                    \li $x_b \coloneqq x_b + 1$
                \End
                \li $x_r \coloneqq x_r + 1$.
            \End
        \end{codebox}
    \end{minipage}%
    \begin{minipage}{.5\linewidth}
    \begin{codebox}
        \Procname{$\proc{StandardMerge}(a[x_a, y_a), b[x_b, y_b), r[x_r, -))$}
            \zi \Comment{Merge $a[x_a, y_a)$ and $b[x_b,
            y_b)$ into $r[x_r, -)$,}
            \zi \Comment{assuming there is enough output space.}
            \li \While $x_a < y_a \land x_b < y_b$ \Do
                \li \If $a[x_a] \leq b[x_b]$ \Then
                    \li $r[x_r] \coloneqq a[x_a]$
                    \li $x_a \coloneqq x_a + 1$
                \li \Else
                    \li $r[x_r] \coloneqq b[x_b]$
                    \li $x_b \coloneqq x_b + 1$
                \End
                \li $x_r \coloneqq x_r + 1$.
            \End
        \end{codebox}
    \end{minipage}
    \caption{Standard and page-based binary straight merging algorithm. The page-based method use one extra check (boxed) that the standard merge doesn't need.}
    \label{alg:page-standard-merge}
\end{algorithm}

Now, a naive page merging algorithm is as in \wref{alg:page-standard-merge}.
We imagine, within
\proc{PageMerge}, that the indices $x_a, y_a, x_b, y_b, x_r, y_r$ are passed
by reference. This algorithm merges two (perhaps partial) pages of memory into a
(perhaps partial) result page. We can merge two buffers by repeatedly calling
this subprocedure. Note that the cost of the subprocedure dominates over the
cost of bookkeeping around it, since we only call such a subprocedure when we
finish processing pages --- this is roughly once for every merge, plus once
every roughly $P$ steps.

%
%
%
%
%
%
%
%
%
%
%
%
%
%
%
%
%
%
%
%

Compare this with a normal merging algorithm, where the output range is assumed
to be long enough to contain all the output we care to write
(cf.~\wref{alg:page-standard-merge}).
Note that \proc{StandardMerge} does one fewer check than
\proc{PageMerge} per element; namely, it doesn't do the check $x_r < y_r$.
But, conveniently the value $y_r - x_r$ decreases by exactly 1 every loop
iteration. Hence we can do loop unrolling to get rid of almost all of these
checks. This modification makes the unrolled \proc{PageMerge} and \proc{StandardMerge}
do the same operations most of the time~-- leading to our almost inplace
algorithm doing almost the same innermost loop as the naive algorithm.

\paragraph{Pre-allocation of buffers}
To reduce the cost of allocating and deallocating memory, as well as to improve
memory locality, it is advantageous to allocate all the extra memory used
by the algorithm once, at the beginning of the run. Simply allocate enough space
for all the memory we could need in the worst case, and add the allocated pages
to the free list to begin with.

\paragraph{Eliminating linked lists}
Practical implementations using linked lists can suffer from (i) higher memory
consumption, and (ii) reduced performance due to more metadata operations
(e.g.~dereferences, allocations, deallocations).
Therefore, we optimise by storing page lists (as well as the free page list) as
arrays. Here one can take advantage of the fact that (other than $O(1)$ page
lists used for \id{curr}, \id{next} and for merging), all other runs are stored
on a stack. Therefore, we can keep a stack of arrays of pointers to pages for
the runs on the stack, where the page pointer arrays for the constituent buffers form subsegments of that one stack --- we also pre-allocate all of the potential stack memory from the beginning of the run of our algorithm.

\paragraph{Short run merges}
If we merge two runs who, together, can fit within a single page, there is no
point in removing a new page from the free list for the answer and placing the
pages of the runs we merge back to the free list. We can slightly optimise this
case by just merging into one of the two already allocated pages.

\section{Evaluation}

We formulate the following hypotheses to test in our empirical evaluation.
Note that the last two hypotheses can be confirmed by mathematical analysis and would hence not require an empirical validation; we use them as sanity checks in our evaluation.

\begin{enumerate}[label=\textsf{\bfseries\color{lipicsGray} (H\arabic*)},ref={(H\arabic*)},leftmargin=3em]
	\item \label{hyp:memory-savings}
		Almost inplace merging allows for a substantial reduction in memory consumption.
	\item \label{hyp:low-overhead}
		Almost inplace merging has low running-time overhead.
	\item \label{hyp:inplace-not-competitive}
		Fully inplace merging (as a blackbox, or weaved into an overall clever algorithm) is not competitive with non-inplace sorting.
	\item \label{hyp:copy-expensive}
		Avoiding extra copies resp.\ moves helps when sorting larger objects.
	\item \label{hyp:comparisons-correct}
		The number of comparisons is unaffected by the run storage method.
	\item \label{hyp:moves-correct}
		Data movement can be predicted accurately based in terms of the mergecost $\mathcal M$.
\end{enumerate}

\paragraph{Experimental setup}
We implemented the variants of Powersort in \texttt{C++} and compared them there with the GNU Compiler Collection (GCC) implementation of \texttt{std::stable\_sort} and \texttt{Wikisort} across different scenarios. We tested three different metrics: the time needed by the algorithm, the number of comparisons used by the algorithm, and the number of assignments used by the algorithm. The full code of our experimental setup is included in the supplemental material mentioned below the abstract. (The code for memory calculation and for properly counting assignments is in a feature branch.)

\paragraph{CPU and compiler settings} We ran our experiments on a                      \texttt{Intel(R) Core(TM) i7-4770 CPU @ 3.40GHz}. We compiled with the \texttt{-O2} flag on \texttt{gcc version 10.2.1 20210110 (Debian 10.2.1-6)}. 
We realised late in testing that we were counting assignments incorrectly (i.e.~not counting copy or move constructors, or move assignment); we retested using the code from branch \texttt{assignment-fix} from the repository.

\paragraph{Algorithm variants} We run 6 different algorithms:
\begin{enumerate}
    \item Virtual-Memory Powersort,%
    \item CPython-style Powersort,%
    \item Pingpong Powersort,%
    \item Powersort with \texttt{C++} standard library inplace merge,%
    \item Wikisort~\cite{McFadden2021},%
    \item \texttt{C++} standard library \texttt{std::stable\_sort}. %
\end{enumerate}
The GCC implementation uses different algorithms depending on whether allocating a buffer of size $n$ is successful, switching between methods in the recursion.
For our running-time experiments, \texttt{std::stable\_sort} was always provided with 
sufficient buffer space, so it executes a Bottom-up Pingpong Merge (with Insertion sort for initial run formation).

For all of our Powersort implementations, we forcibly pre-sort short runs using insertion sort, up to a certain minimum run length. This minimum run length is chosen according to Timsort's rule: to compute this minimum run length, start from $\ell = n$, then repeatedly set $\ell = \lceil \ell / 2 \rceil$ until $\ell < 64$. Then $\ell$ is the required minimum run length.
In this work, we consider only increasing runs; all Powersort variants could take advantage of decreasing runs, as well.

\paragraph{Input data types}

We next describe \emph{what} we sort.
Observe that a data type is (primarily) characterised by two properties: the amount of memory it consumes (and thus requires to move), and the efficiency of pairwise comparisons. We therefore distinguish four distinct scenarios depending on whether comparison resp.\ move time is high resp.\ low.
\begin{description}
    \item[Fast comparison, fast move.]
    For this scenario, we use 32 bit integers. In our graphs, this data type is called \texttt{int}.
    \item[Slow comparison, fast move.] 
    For this scenario, we use pointers to arrays of length 30, containing 32-bit integers, ordered lexicographically. For this scenario, the first 29 integers are always set to 0, so on top of the cost of pointer dereferencing, each comparison compares all 30 pairs of integers, making comparisons rather costly. In our graphs this data type is called \texttt{ptr}.
    \item[Fast comparison, slow move.] 
    For this scenario, we use arrays of length 30 containing random 32-bit integers, ordered lexicographically. Here, all elements are potentially nonzero, so most comparisons will only compare one pairs of integers. In our graphs, this data type is called \texttt{randomBlob30}.
    \item[Slow comparison, slow move.] 
    For this scenario, we use arrays of length 30, containing 32-bit integers, ordered lexicographically, with only the last integer nonzero.
    In our graphs, this data type is called \texttt{zeroBlob30}.
\end{description}
For investigating running time and memory, we used all 4 data types. 
(For counting comparisons and assignments, the underlying data type is irrelevant; we use \texttt{int} for efficiency.)

\paragraph{Input permutations}
All our randomness is generated using the \texttt{mt19937} \texttt{C++} standard library pseudorandom generator, with a fixed seed (hence, the inputs we run our algorithms on are the same between all algorithms).
Let $N = 10^7$. We sort input sequences whose length is distributed uniformly at random between $9N/10$ and $N$. After having fixed the run length, we generate a random input sequence. Depending on the data type, we generate either single integers or arrays with either some values equal to 0, or all values random, as described above. Every integer we want to generate randomly is taken uniformly at random from the range $[100, 10^9]$. After generating a random input sequence, we introduce pre-sortedness into it in the following way; let $S \in \mathbb{N}$ be a parameter we fix. We repeatedly draw a geometrically distributed length and pre-sort a further run with that length. The geometric distribution has parameter $1 / S$ --- this means that the expected run length is $S - 1$, i.e.~we expect to have $S$ consecutive $<$ signs from the beginning of each run (ignoring the possibility that two consecutive runs happen to accidentally form a single run).

We tested our algorithms with $S \in 
\{2, 10^2, 10^3, 10^4, 10^5, 10^6\}$. We only considered one case with small $S$, since our algorithms use insertion sort to guarantee a minimum run length.

\paragraph{Number of iterations} For each combination of data type, algorithm, presortedness and metric, we ran 100 experiments and aggregated the results. 

\paragraph{Memory consumption} Additionally to the experiments mentioned before, we computed the memory consumption (excluding stack space and constantly many local variables) for the CPython style implementation of Powersort, the few-moves implementation of Powersort, and the almost-inplace implementation of Powersort. Since this does not depend on the average run-length, we only computed it for $S = 2$.

\section{Results}

We consider in turn the data relevant for our hypotheses from above.
\wref{fig:memory} compares the extra memory allocated by the algorithms.
\wref{fig:time} and \wref{fig:time-no-inplace} show the running-time results.
\wref{fig:assignments} and \wref{fig:comparisons} give the combinatorial cost measures (comparisons resp.\ assignments).

\begin{figure}[tbh]
	\centering
	\def\mathdefault#1{#1}
	\resizebox{0.75\linewidth}{!}{\input{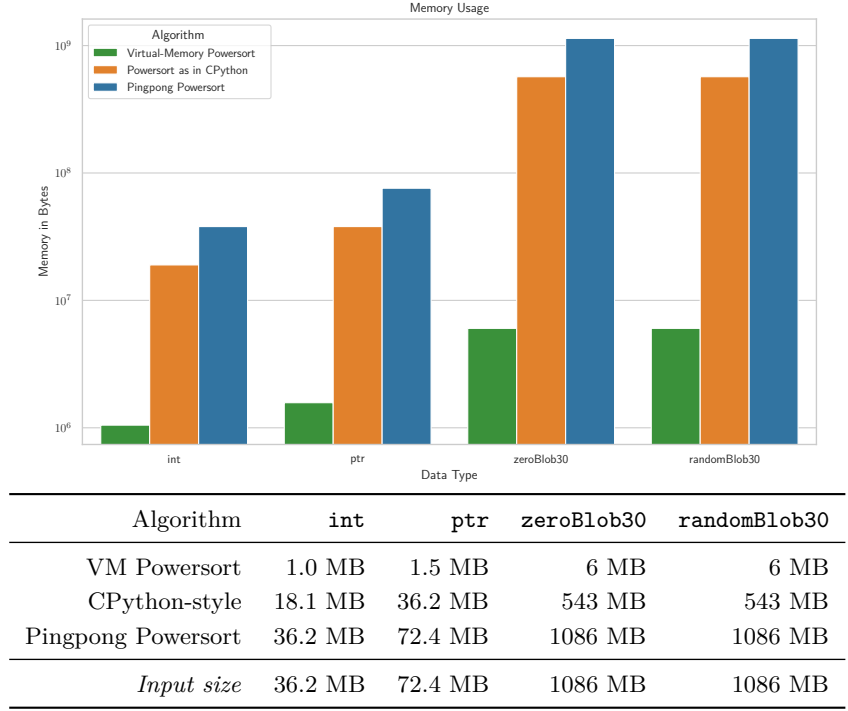}}
	\begin{tabular}{rrrrr}
	\toprule
		          Algorithm & \texttt{int} & \texttt{ptr} & \texttt{zeroBlob30} & \texttt{randomBlob30} \\
	\midrule
		       VM Powersort &       1.0 MB &       1.5 MB &                6 MB &                  6 MB \\
		      CPython-style &      18.1 MB &      36.2 MB &              543 MB &                543 MB \\
		 Pingpong Powersort &      36.2 MB &      72.4 MB &             1086 MB &               1086 MB \\
	\midrule
		\textit{Input size} &      36.2 MB &      72.4 MB &             1086 MB &               1086 MB \\
	\bottomrule
		                    &              &              &                     &
	\end{tabular}
	\caption{Average memory usage in bytes of Virtual-Memory Powersort, Powersort as in CPython and Pingpong Powersort, for the different input types. This counts only main buffer allocations, not $O(\log(n))$-sized buffer stacks or $O(1)$ local variables. \emph{The $y$-axis is logarithmic!}}
	\label{fig:memory}
\end{figure}

\begin{figure}[tbh]
    \centering
    \resizebox{1\linewidth}{!}{\input{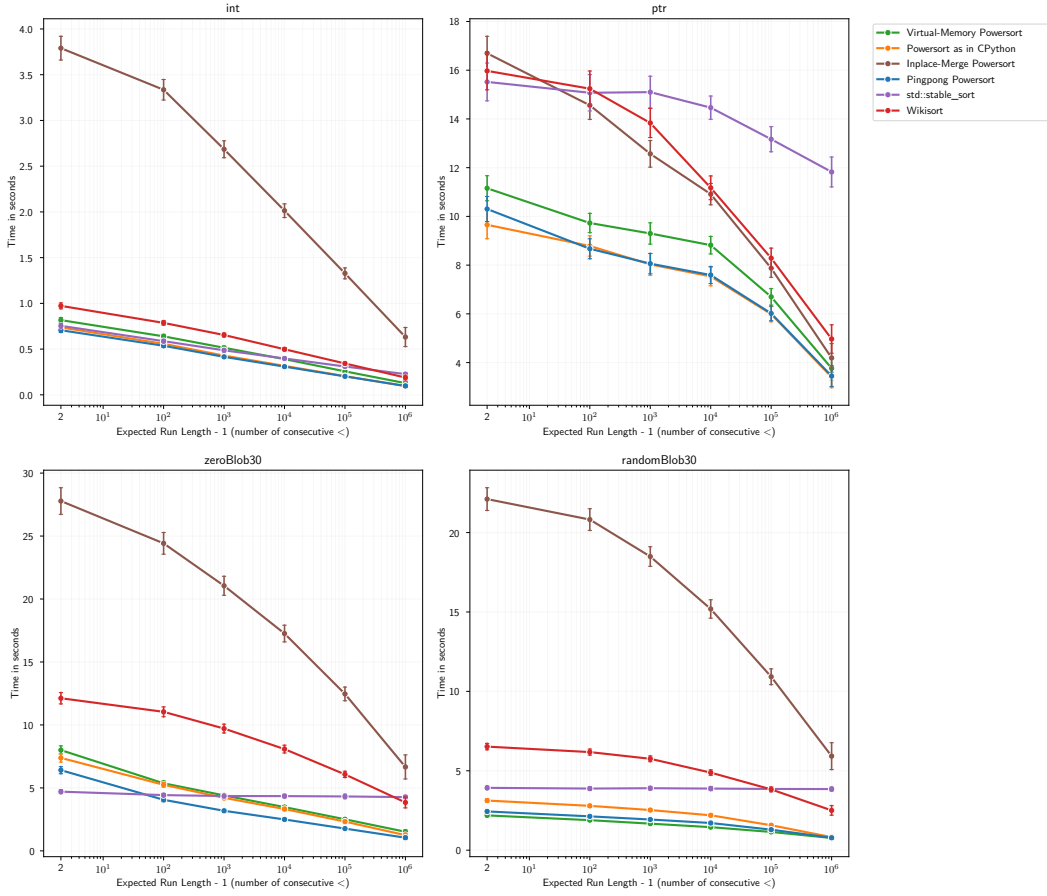}}
    \caption{Average running time in seconds for the different input types. The input size is uniformly chosen between 9 million and 10 million, and the $x$-axis varies the presortedness.}
    \label{fig:time}
\end{figure}

\begin{figure}[tbh]
    \centering
    \resizebox{1\linewidth}{!}{\input{plots/plot_time_errorbars1.pgf}}
    \caption{Same as \wref{fig:time} but without the Inplace-Merge Powersort.}
    \label{fig:time-no-inplace}
\end{figure}

\begin{figure}[tbhp]
    \centering
    \resizebox{1\linewidth}{!}{\input{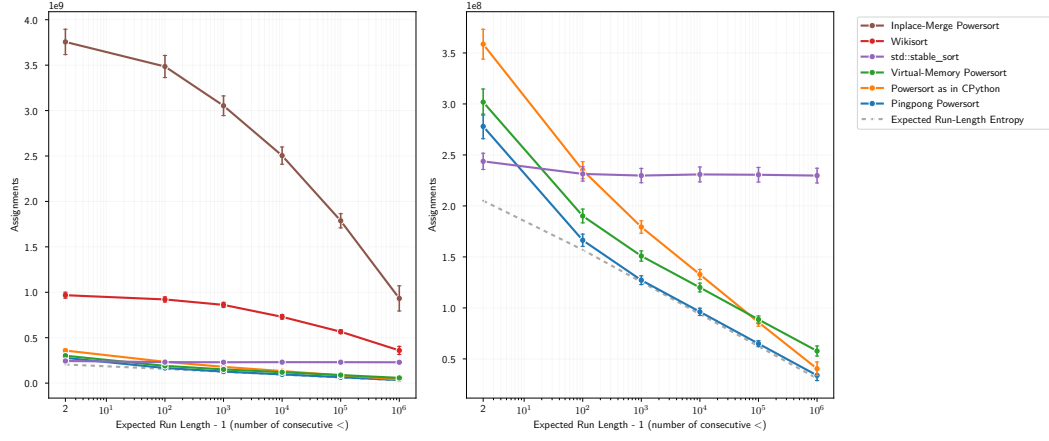}}
    \caption{The average number of move/copy assignments and move/copy constructions for the same inputs as in \wref{fig:time} with all algorithms (left) and without Wikisort and Inplace-Merge Powersort (right).
    The expected run-length entropy is calculated by the formula $\textit{EN} \log_2(\textit{EN} / (S + 1))$, where $\textit{EN} = 0.95 \times 10^7$ is the expected input length, and $S + 1$ is the expected run length (ignoring consecutive runs that happen to become one run).
    }
    \label{fig:assignments}
\end{figure}

\begin{figure}[tbhp]
    \centering
    \resizebox{0.66\linewidth}{!}{\input{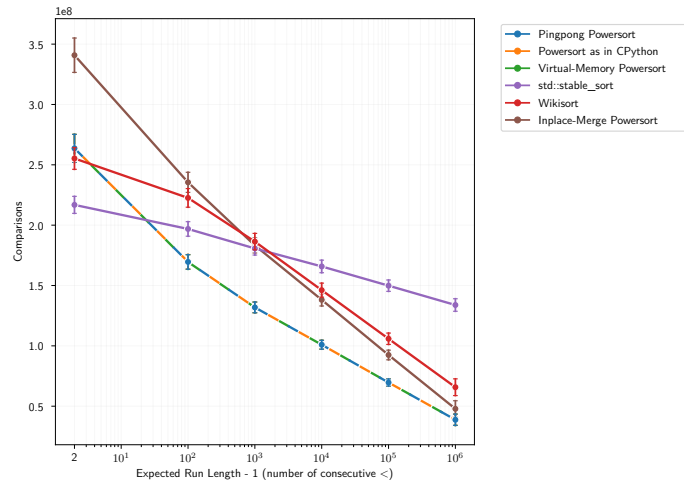}}
    \caption{
    	The average number of comparisons for the same inputs as in \wref{fig:time}.
    	Pingpong, CPython-style and VM Powersort, execute (almost) the same comparisons.
    }
    \label{fig:comparisons}
\end{figure}

\paragraph{\wref{hyp:memory-savings}: Significant reduction in memory consumption}

As described, the required buffer area is reduced from space for $n/2$ objects to $O\left(\sqrt{n \log n}\right)$ objects, where $n$ represents the number of objects to be sorted. The experimental results in \wref{fig:memory} show the reduction in memory:
Almost-inplace Powersort achieves a memory reduction of several orders of magnitude compared to the CPython-style implementation and the few-moves implementation of Powersort.

\paragraph{\wref{hyp:low-overhead}: Almost inplace merging has low overhead} 

We compare the run-time of almost-inplace Powersort with the other algorithms. We see in \wref{fig:time} and \wref{fig:time-no-inplace} that almost-inplace Powersort performs comparably with previous implementations of Powersort, and is uniformly faster than \texttt{Wikisort} or a black-box inplace implementation of inplace Powersort. 
Of particular note is that for \texttt{randomBlob30}, almost-inplace Powersort is even faster than the other algorithms. This might be due to touching an overall smaller range of memory addresses; here, the input already occupies 1GB of memory.

\paragraph{\wref{hyp:inplace-not-competitive}: Fully inplace merging is not competitive} 

We see, in \wref{fig:time}, that an implementation of Powersort that uses a \texttt{C++} library inplace merge to achieve lower memory consumption is incredibly slow. 
This is probably due to the large number of assignments shown in \wref{fig:assignments}. 
Therefore, more clever techniques are needed. 
Wikisort, which also uses inplace merging as a subroutine, makes progress on this front,
however it isn't fully run-adaptive (as seen by the higher comparison count, \wref{fig:comparisons}),
and still significantly slower.

\paragraph{\wref{hyp:copy-expensive}: Avoiding extra copies resp.\ moves helps when sorting larger objects} 
Moving elements is very cheap for \texttt{int} and \texttt{ptr} data, but has substantial cost
for \texttt{zeroBlob30} and \texttt{randomBlob30}.
Comparing the relative ranking of the algorithms in \wref{fig:time-no-inplace} across input types,
we see that the move-heavy CPython-style Powersort and the move-optimized Pingpong Powersort perform
almost the same for small types, whereas the former is substantially worse on large objects.
Note that \texttt{std::stable\_sort} on random inputs is slightly faster on \texttt{zeroBlob30}.
This is likely due to its lower comparison cost: 
the adaptive algorithms' method for generating initial runs is not optimized for very expensive comparisons.

\paragraph{\wref[Hypotheses]{hyp:comparisons-correct} and~\ref{hyp:moves-correct}: Theoretical move and comparison counts} 

Our results for counting moves and comparisons closely match the theoretical estimation.
Recall that VM Powersort should use $\mathcal{M} + 3n$ moves, Pingpong Powersort should use $\mathcal{M} + n$ moves, and the CPython-style Powersort should use $1.5\mathcal{M} + n$ moves; 
all of these should use $\mathcal{M} + n$ comparisons. 
We see this is well predicts the actual numbers, when ignoring $S = 2$ (where the initial run formation has a large influence). 
In particular, we see how closely the comparisons coincide in \wref{fig:comparisons}.\footnote{Technically there is a very small difference, since some of the algorithms do some of their merges left-to-right or right-to-left respectively. This difference is very small.} 
In \wref{fig:assignments}, we see how (in our logarithmic plot) the trends for the three algorithms are clearly linear (which we expect, since for fixed $n$, $\mathcal{M}$ is logarithmic with respect to inverse run lengths), with slopes as expected (see the expected run-length entropy, in gray). The only unexpected thing is that Pingpong Powersort performs better than expected --- likely since the optimisation mentioned in the last paragraph of \wref{sec:fewMoves} occurs often in our random inputs.

\section{Conclusion}
\label{sec:conclusion}

We conclude from the support of our hypotheses that in \texttt{C++} generic sorting, 
Pingpong Powersort is a strong contender for a good \texttt{std::stable\_sort} implementation
when ample extra buffer space is available. 
Only one out of many scenarios favors the current Pingpong Bottom-up Mergesort:
if comparisons are expensive, objects are big, \emph{and} the input is a random permutation.
In the context of adaptive sorting, we point out that Pingpong Bottom-up Mergesort
can save some of the comparisons when natural runs exist, but its memory movement is not adaptive
and hence it quickly falls behind when sorting larger objects.

Moreover, almost-inplace Powersort serves a wide middle ground where substantially less, but still some
buffer space is acceptable. Its slowdown over the linear-buffer variants is very modest,
but its quadratically smaller memory footprint makes fast, optimally-adaptive, stable sorting applicable
to a much wider range of applications.

	\myacknowledgements
%

%
\bibliography{references}

\clearpage
\appendix
\ifkoma{\addpart{Appendix}}{}

\ifdraft{
	\clearpage
	\part*{Notes-to-self}
	\printnotestoself
}{}

\end{document}